\documentclass[]{spie}  

\usepackage[]{graphicx}

\title{URAT: astrometric requirements and design history} 

\author{Norbert Zacharias\supit{a}, Uwe Laux\supit{b},
        Andrew Rakich\supit{c}, and Harland Epps\supit{d}
\skiplinehalf
\supit{a}U.S.~Naval Observatory, 3450 Mass.~Ave.~NW, Washington DC 20392\\
\supit{b}Th\"uringer Landessternwarte, Sternwarte 5,  07778 Tautenburg, Germany\\
\supit{c}EOS Space Systems, 111 Canberra Ave, Griffith, ACT 2603, Australia\\
\supit{d}UC, Santa Cruz, Lick Observatory, Nat.Sc.2, Rm 191, Santa Cruz, CA 95064
}

\authorinfo{Further author information: (Send correspondence to N.Z.)\\
  N.Z.: E-mail: nz@usno.navy.mil, telephone: 1 202 762 1423\\
  U.L.: E-mail: laux@tls-tautenburg.de \\
  A.R.: E-mail: arakich@eos-aus.com \\
  H.E.: E-mail: epps@skewray.ucolick.org}

\begin{document} 
  \maketitle 

\begin{abstract}
The U.S.~Naval Observatory Robotic Astrometric Telescope (URAT)
project aims at a highly accurate (5 mas), ground-based, all-sky
survey.
Requirements are presented for the optics and telescope for this
0.85 m aperture, 4.5 degree diameter field-of-view, specialized
instrument, which are close to the capability of the industry.
The history of the design process is presented as well as
astrometric performance evaluations of the toleranced, 
optical design, with expected wavefront errors included.
\end{abstract}


\section{INTRODUCTION}

From 1998 to 2004 the U.S.~Naval Observatory (USNO) conducted an
astrometric, all-sky survey to 16th magnitude with its 20 cm aperture
Twin Astrograph, resulting in the USNO CCD Astrograph Catalog (UCAC),
with its second release (UCAC2) made public in 2003 \cite{ucac2}.
Around the year 2000 the USNO began preparations for a follow-up project,
the USNO Robotic Astrometric Telescope (URAT), a dedicated, 
astrometric, 1-meter size telescope with a wide field of view for
an all-sky survey going much deeper than UCAC and on a positional
accuracy level aiming at 5 milliarcsecond (mas) standard error per 
coordinate.

The idea of such a telescope goes back to the late 1980s \cite{cdv89},
then envisioned for photographic plates.
The purpose of such a telescope is threefold. 
First URAT aims at the densification of the
celestial reference frame, providing a large number of accurate reference 
stars \cite{a-surveys} to support, for example, general, deep sky surveys
like PanSTARRs and LSST, and to support astrometry in our solar system,
as outlined in the recent decadal survey \cite{decadal}.
Second, URAT provides absolute proper motions on an inertial system,
enabling significant galactic dynamics studies in the pre-Gaia era.
Third, URAT will be able to observe trigonometric parallaxes of
many thousands of stars unbiased by selection effects like
high proper motion targets.  More details about the URAT project 
in general can be found elsewhere \cite{potsdam,lowell1}.

\section{REQUIREMENTS}

\subsection{General} 

The basic parameters and requirements of the proposed new telescope
are summarized in Table 1.

\begin{table}[h]
\caption{Basic data of URAT.} 
\label{tab:basic}
\begin{center}       
\begin{tabular}{|l|rll|} 
\hline
\rule[-1ex]{0pt}{3.5ex}  aperture           & 0.85 & m & \\
\rule[-1ex]{0pt}{3.5ex}  effective aperture & 0.60 & m & equivalent unobstr.~area \\
\rule[-1ex]{0pt}{3.5ex}  focal length   & 3.60 & m & \\
\rule[-1ex]{0pt}{3.5ex}  image scale    & 57.3 & "/mm & = 57.3 mas/$\mu$m \\
\hline
\rule[-1ex]{0pt}{3.5ex}  field of view  & 4.00  & deg  & diameter design goal \\
\rule[-1ex]{0pt}{3.5ex}                 & 4.50  & deg  & usable (vignetted)\\ 
\rule[-1ex]{0pt}{3.5ex}  passband       & 650 $-$ 800  & nm& astrometric \\
\rule[-1ex]{0pt}{3.5ex}                 & 500 $-$ 950  & nm& photometric \\
\rule[-1ex]{0pt}{3.5ex}  stray light, ghost images & \multicolumn{3}{l|}
                   {10 magnitudes fainter than direct light / surface area}\\
\hline
\rule[-1ex]{0pt}{3.5ex}  focusing & \multicolumn{3}{l|}
                           {by moving the backend of the telescope} \\
\rule[-1ex]{0pt}{3.5ex}  telescope flip:& \multicolumn{3}{l|}
                   {2 telescope orientations, $180^{\circ}$ w.r.t.~sky} \\
\rule[-1ex]{0pt}{3.5ex}  mount & \multicolumn{3}{l|}
                           {equatorial} \\
\rule[-1ex]{0pt}{3.5ex}  guiding & \multicolumn{3}{l|}
                           {active, with 2 guide detectors in focal plane} \\
\rule[-1ex]{0pt}{3.5ex}  operation & \multicolumn{3}{l|}
                           {robotic} \\
\rule[-1ex]{0pt}{3.5ex}  detector  & \multicolumn{3}{l|}
                           {4 CCDs, each 10.6k $\times$ 10.6k pixel} \\
\hline
\end{tabular}
\end{center}
\end{table} 

The goal is to reach about magnitude 21 in a few minutes exposure
time and cover an entire hemisphere about 6 times per year
in a relatively narrow astrometric passband.
In order to cover a large amount of sky, critical sampling 
near 2 pixel per full width at half maximum (FWHM) is required. 
The limiting magnitude requirement leads to a 1-meter class aperture.
In order to ease the f/ratio requirement, a small pixel size 
(about 9 $\mu$m) was adopted.
These basic project parameters then determine the focal length
and field-of-view requirements.

The passband needs to be in the V to I range (peak detector quantum
efficiencies, large flux of ``normal" stars).
Exclusion of the $H_{\alpha}$ line is important to avoid 
complications with galactic nebulae; the emphasis is on stars.
The UCAC project chose a bandbass just blue of $H_{\alpha}$,
while for URAT a passband just red of $H_{\alpha}$ has been adopted.
This will mitigate the differential refraction effects, both from
the Earth's atmosphere and from refractive optics design considerations.
The width of the passband is chosen as a compromise between reaching
a deep limiting magnitude and minimizing differential color effects.
The optics design goal was set to cover 650 to 800 nm, while the
passband to be used will likely be about 670 to 770 nm.

\begin{table}[hb]
\caption{Specific astrometric requirements for URAT.} 
\label{tab:astrom}
\begin{center}       
\begin{tabular}{|l|l|} 
\hline
\rule[-1ex]{0pt}{3.5ex}  general:  & \\
\rule[-1ex]{0pt}{3.5ex}  good image quality  & $\ge$ 70 \% Strehl ratio \\
\rule[-1ex]{0pt}{3.5ex}  entrance pupil & circular symmetric, no spider structure \\
\rule[-1ex]{0pt}{3.5ex}  flat field      & no significant field curvature\\
\rule[-1ex]{0pt}{3.5ex}  stability       & variations of ``high order terms" \\
\rule[-1ex]{0pt}{3.5ex}                  & on $\le$ 100 nm level (thermal, flexure)\\
\hline
\rule[-1ex]{0pt}{3.5ex}  distortion: & \\
\rule[-1ex]{0pt}{3.5ex}  geom.~optical distortion  & $\le$ 1 arcsec / degree$^{3}$ \\
\rule[-1ex]{0pt}{3.5ex}  deviation from 3rd order dist.& $\le$ 300 nm over $4^{\circ}$
                                                               diameter field \\
\hline
\rule[-1ex]{0pt}{3.5ex}  lateral color: & \\
\rule[-1ex]{0pt}{3.5ex}  center-of-mass centroids & within 400 nm over astrom.~band \\
\rule[-1ex]{0pt}{3.5ex}  peak location  centroids & within 400 nm over astrom.~band \\
\hline
\rule[-1ex]{0pt}{3.5ex}  symmetric images: & \\
\rule[-1ex]{0pt}{3.5ex}  all PSFs + 0.7" seeing & centroids 90\% and 1\% within 400 nm \\
\hline
\end{tabular}
\end{center}
\end{table}

The ``flip" of the telescope is a required to perform astrometric
calibration observations.  The same area of the sky will be
observed with the entire telescope tube assembly plus camera
in 2 possible orientations which are rotated by $180^{\circ}$
around the optical axis with respect to each other. 
At the astrograph we achieved this by observing with the 
telescope from the East and West side of the pier (B\&C equatorial mount).
An equatorial fork mount would be an option for URAT if it
can be rotated by a sufficiently large angle and if the telescope
can swing through the ``fork" to both sides.

A requirement which entered the process at a late stage was
the ability of URAT to also perform photometric surveys in at
least 2 colors within the V to I range of the spectrum.
Color information is required for many applications to predict
the brightness of stars in non-standard passbands often used
in DoD instrumentation.
No strict astrometric performance is required for the 
500 to 900 nm band; a general ``good" image quality, like
``nearly diffraction limited" for the expected medium
seeing of 1 arcsec FWHM is sufficient for photometric
surveys, with re-focusing of the instrument between filters 
allowed.
The detector development is in progress\cite{sta} with a funded phase II
Small Business Innovation in Research (SBIR) program.
A complete prototype camera including a single 10.6k by 10.6k
thinned CCD is expected to be delivered to USNO by the end of 2006\cite{mike}.

\subsection{Astrometric Requirements}

The requirements specifically relevant for astometry are summarized
in Table 2.
The general image quality, which is typically defined by a Strehl ratio
is often the most important requirement for telescope designs, demanding
a nearly diffraction limited design over a specified field of view and passband.
For URAT this is of minor importance and generally ``sharp" images follow from
the other, more specific requirements for astrometric mapping.

Particularly problematic is the requirement of a circular symmetric
entrance pupil without secondary mirror spider structures which 
prohibited the use of most traditional optical designs.
The reason for this requirement is to be able to measure very bright
and very faint stars simultaneously.  It is assumed that the detector
for URAT does not ``bleed" charge when saturation is reached.
Accurate positions could then be obtained even from overexposed
stellar images by just rejecting the central, saturated pixels of the
image profile.  Conventional designs with diffraction spikes from 
secondary mirror support structures would prevent this.
Curved spiders were suggested \cite{spider1,spider2} which reduce the effect 
of ``spikes", but a rigorous approach was adopted here particularly for
DoD applications.

At first it seems that a general pattern of geometric optical distortion, 
even if very large, could be tolerated for an astrometric telescope as long
as the stability is guaranteed.  However, extensive calibrations would be
required to achieve satisfactory astrometric results from such a telescope,
which is rendered impossible in practice when considering the limited number
of reference stars available with typically unknown colors or considering
the number of ``bins" needed to map out the focal plane geometric distortions 
to within a specified, tight tolerance and bridge gradients between bins.
Thus a relatively small optical distortion is required to begin with,
which then will need to be calibrated to the measure accuracy level
with as few parameters as possible.  That is why only a 3rd order term
is allowed here with higher order deviations smaller than 300 nm 
in the focal plane, which corresponds to 17 mas in extreme cases 
(usually at the edge of the field), with much better performance on average.

Because of the generally unknown spectral energy distribution of the stars 
to be observed, the images of stars must have ``the same" centroid regardless
of wavelength, which leads to a strict lateral color requirement.
For all field points within the $4^{\circ}$ diameter field of view
and for all wavelengths in the astrometric passband (650--800 nm)
the maximal difference in image centroids of any 2 monochromatic 
point spread functions (PSF) shall not exceed 400 nm for a 90\% confidence
level of the ``as-built" system.\footnote{The wording ``as-built"
used in quotes is used in this paper for an optical system which includes
expected wave-front errors from manufacturing and alignment errors
as derived from simulations of the toleranced system.}
When dealing with such tight tolerances the ``centroid" needs to
be defined more precisely.
Due to possible small image profile asymmetries (coma aberrations)
there are different definitions of ``centroid".
Here the lateral color requirement needs to be obeyed by 2 definitions:
center-of-mass centroid of a PSF (taking the entire flux) and the
peak location of the PSF, as defined by the center-of-mass
position of the 90\% and above contour level of the PSF, after
folding with a 2-dimensional Gaussian seeing profile of 0.7 arcsec FWHM.
The first lateral color criterion could be checked by ZEMAX
directly, while the latter required a dedicated astrometric
evaluation (see below).

\begin{figure}
\begin{center}
\includegraphics[height=6cm]{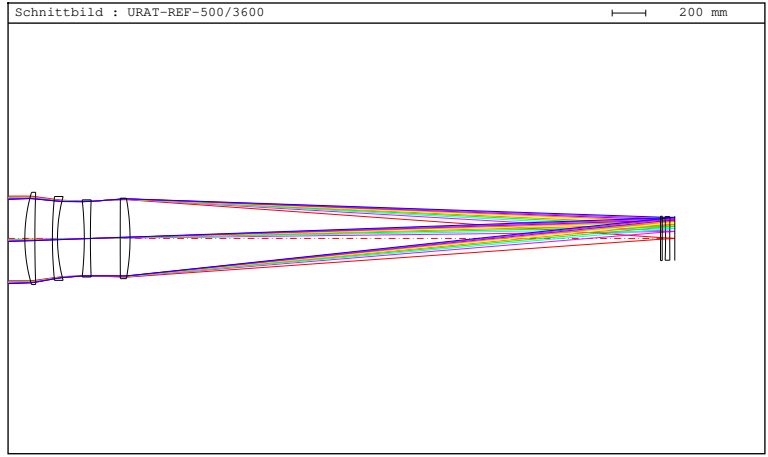}
\end{center}
\caption[ex1] {Optics layout of pure refractive solution.}
\end{figure} 

\begin{figure}
\begin{center}
\includegraphics[height=6cm]{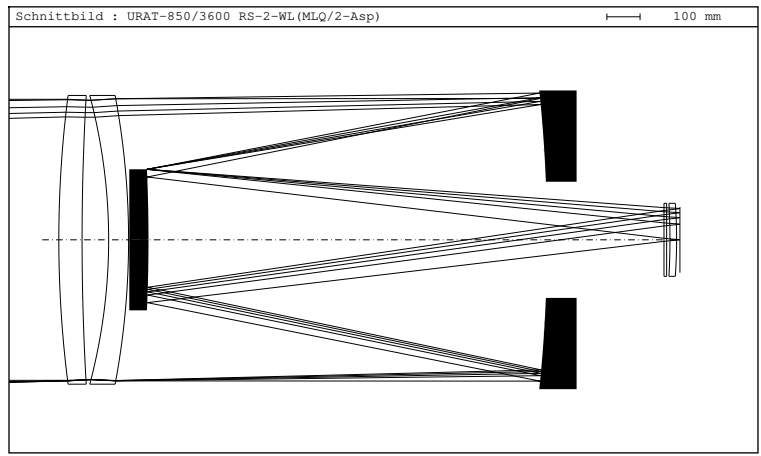}
\end{center}
\caption[ex2] {Optics layout of RS2b design.}
\end{figure} 

\begin{figure}
\begin{center}
\includegraphics[height=6cm]{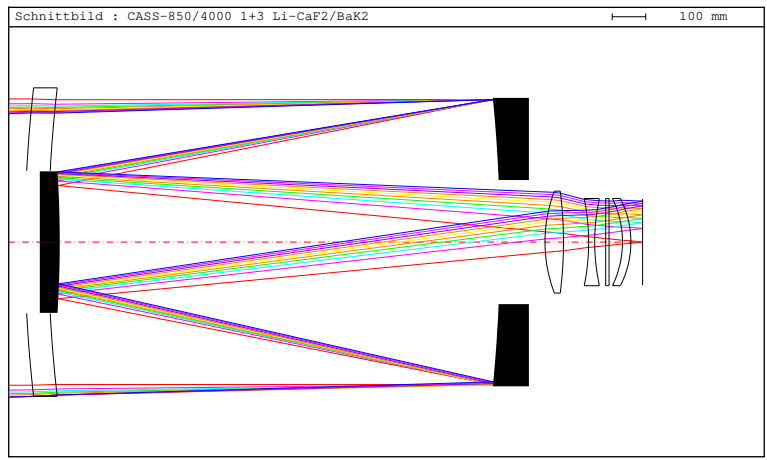}
\end{center}
\caption[ex3] {Optics layout of 1CaF design.}
\end{figure} 

\begin{figure}
\begin{center}
\includegraphics[height=6cm]{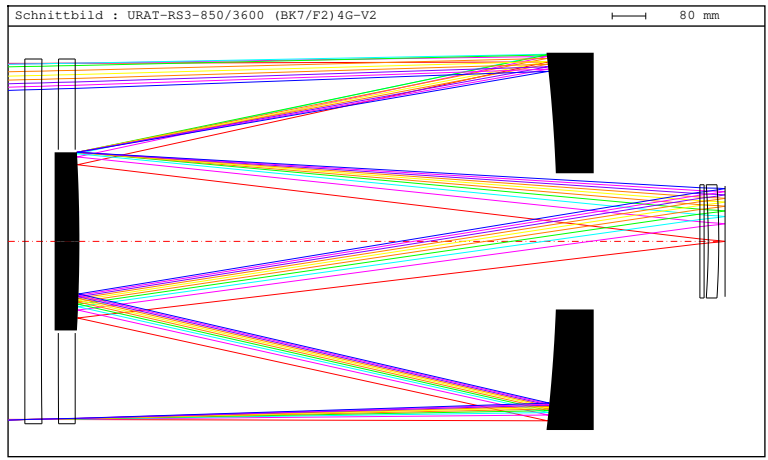}
\end{center}
\caption[ex4] {Optics layout of RS3 design.}
\end{figure} 

\begin{figure}
\begin{center}
\includegraphics[height=6cm]{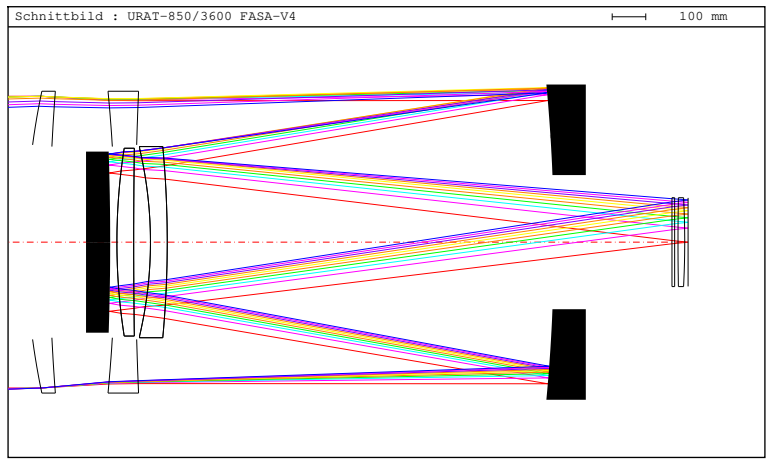}
\end{center}
\caption[ex5] {Optics layout of FASA2 v4 design.}
\end{figure} 

Similarly, the image asymmetry is bound to give a difference in centroid 
positions nowhere exceeding 400 nm for the ``as-built" system and folded
with 0.7 arcsec seeing when comparing the 90\% and 1\% flux level of any PSF
in the astrometric field-of-view and passband.
At 0.7 arcsec seeing the PSFs will be undersampled with about 1.5 pixels
per FWHM and ``pixel-phase" centroid bias effects will be taken into
consideration similarly to the UCAC reduction procedures.\cite{ucac2}

\begin{figure}[ht]
\begin{center}
\begin{tabular}{c}
\includegraphics[height=14cm]{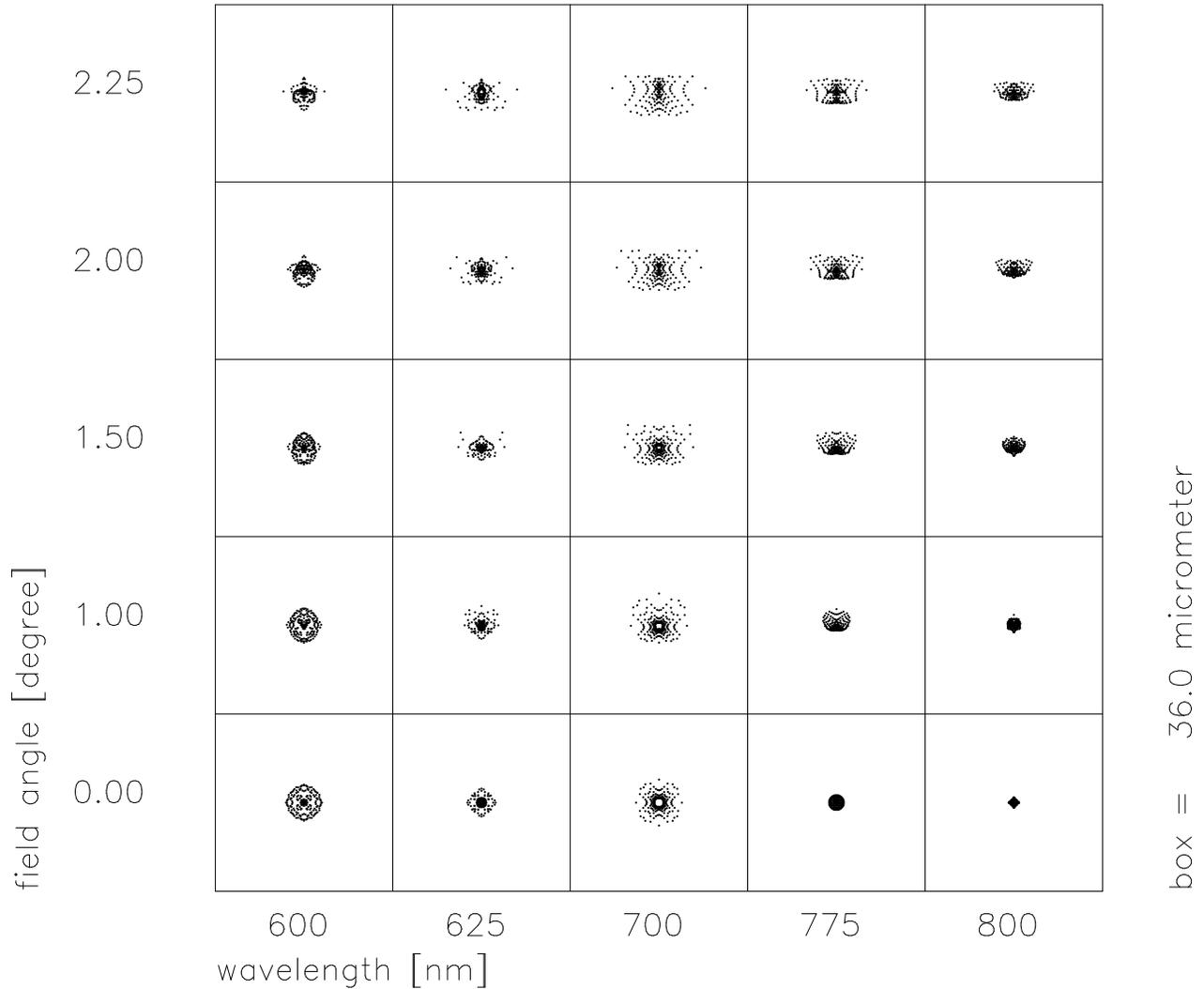}
\end{tabular}
\end{center}
\caption[example] {Spot diagrams of RS3 design.}
\end{figure} 

\begin{figure}[h]
\begin{center}
\begin{tabular}{c}
\includegraphics[height=14cm]{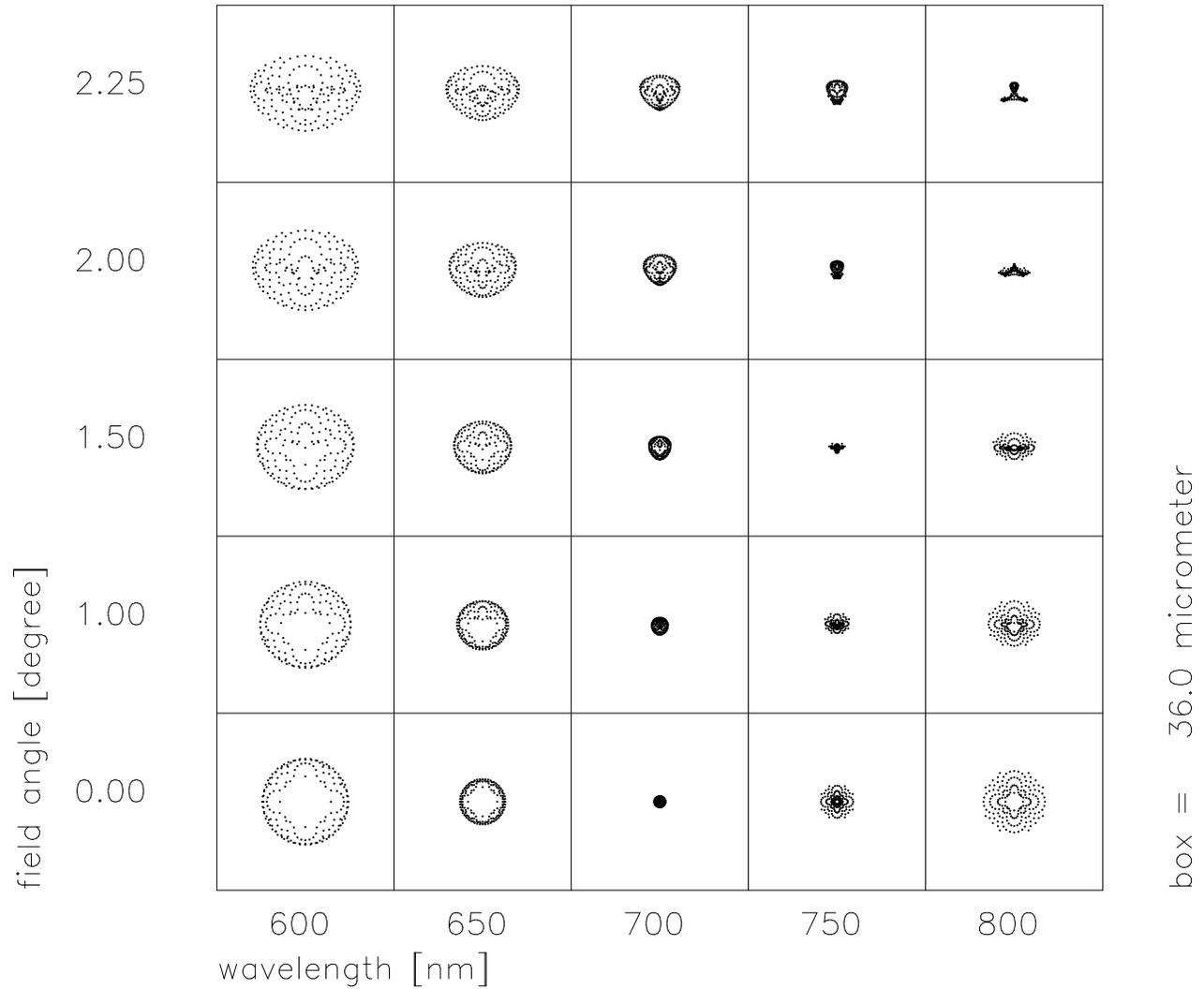}
\end{tabular}
\end{center}
\caption[example] {Spot diagrams of FASA2 v4 design.}
\end{figure}

\section{Design Options}

Design work specifically for URAT began around the year 2001.
Modified Richter-Slevogt systems were proposed by one of us (U.L.),
which were presented e.g.~at the Lowell astrometry meeting
\cite{lowell2}
followed by a single full-aperture lens approach from EOST (A.R.).
Initial contacts with various vendors seemed to indicate that
a 1-meter-class telescope even with ``corrector plates" is not
a problem. 
However, after having presented details (highly aspheric RS3 system 
for example) and the required tolerances, we quickly reached the 
state-of-the-art engineering limits with high price tags.

The 5 designs investigated are summarized in Table 3
with optical layout diagrams shown in Figures 1 to 5.
All but 1 design include only 2 optical elements close to the focal
plane: a filter and dewar window, which are given as ``+2" on
the ``number of elements" line in Table 3.

\begin{table}[h]
\caption{Comparison of initial optical design options for URAT.} 
\label{tab:comp}
\begin{center}       
\begin{tabular}{|l|ccccc|} 
\hline
\rule[-1ex]{0pt}{3.5ex}                    &  refractive  &  RS2b   &1CaF & RS3 & FASA2 \\
\hline
\rule[-1ex]{0pt}{3.5ex} number of elements &      4 + 2   &   4 + 2 &5 + 2& 4 + 2 & 6 + 2\\
\rule[-1ex]{0pt}{3.5ex} number of mirror aspheres&  0     &   1     & 2   &  2   &    1  \\
\rule[-1ex]{0pt}{3.5ex} number of lens   aspheres&  1     &   1     & 1   &  2   &    1  \\
\rule[-1ex]{0pt}{3.5ex} required glass     & special& fus.~sil.&CaF  &BK7, F2&fus.~sil.\\
\hline
\end{tabular}
\end{center}
\end{table}

Due to the large central obstruction of the catadioptric designs
(about 50\% in area) an all-refractive, classical astrograph lens
is an option, giving an equivalent throughput with an aperture of 
about 60 cm.  
The advantages and disadvantages of the individual design options are:

\begin{description}
\item[All-refractive.] PRO:
  \begin{itemize}
    \item no central obstruction
  \end{itemize}
   CON:
  \begin{itemize}
    \item narrow passband (670--750 nm)
    \item need folding to fit in existing domes, asymmetric mechanical setup
    \item availability of optical glass with desired specs is questionable
  \end{itemize}

\item[RS2b.] PRO:
  \begin{itemize}
    \item spherical primary mirror
    \item only fused silica for all refractive elements
  \end{itemize}
   CON:
  \begin{itemize}
    \item 2 full-aperture lenses with 1 aspheric surface
    \item image symmetry not meeting requirements (residual coma)
  \end{itemize}

\item[1CaF.] PRO:
  \begin{itemize}
    \item only 1 full-aperture lens, spherical
  \end{itemize}
   CON:
  \begin{itemize}
     \item issues with CaF lens (thermal and refractive gradients)
     \item limited field-of-view, color correction
  \end{itemize}

\item[RS3.] PRO:
  \begin{itemize}
    \item few elements, smallest amount of glass
  \end{itemize}
   CON:
  \begin{itemize}
    \item 2 full-aperture corrector plates 
    \item 2 highly aspheric surfaces on corrector plates
    \item F2 optical glass for corrector plate may be problematic
  \end{itemize}

\item[FASA2.] PRO:
  \begin{itemize}
    \item only fused silica for all refractive elements
    \item only 1 refractive asphere on sub-aperture lens 
    \item spherical secondary mirror
    \item nearly no geometric optical distortion at all
  \end{itemize}
   CON:
  \begin{itemize}
    \item number of surfaces is large, sub-aperture lenses double-pass
    \item largest central obstruction of presented design options
  \end{itemize}
\end{description}

\section{Selection of Design}

The all-refractive option was discarded when the requirement for
a photometric survey option arose.
The RS2b design has a residual amount of higher order coma which
gives rise to stellar image profile asymmetries exceeding the
strict requirements set for URAT.
The 1CaF design did not meet the later requirements for a larger
field of view, has larger color errors than RS3 or FASA2, and
the CaF lens is problematic due to the large gradients in
thermal expansion and refractive properties combined with the
alignment tolerances and stability requirements.
The RS3 and FASA2 designs have almost equivalent optical performance.
Spot diagrams of the ideal, as designed RS3 and FASA2 systems
are given in Figures 6 and 7, respectively.

However the RS3 design has 2 highly aspherical surfaces on the 
corrector plates and requires BK7 and F2 (or similar) optical glass.
By the end of 2005 the FASA2 design \cite{fasa2} was adopted as URAT
baseline due to easier manufacturability as compared to the RS3 
design.

\begin{figure}
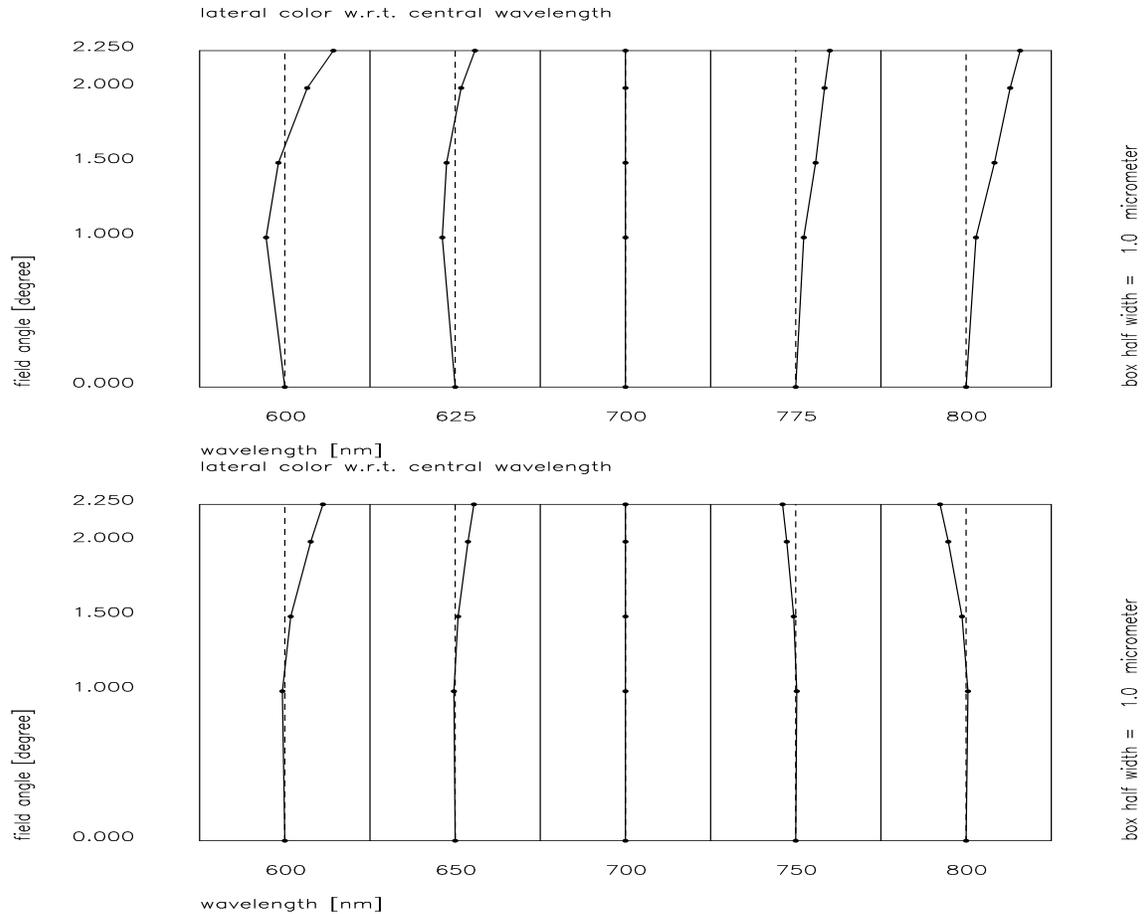

\begin{center}
\includegraphics[width=6.0cm,height=15cm,angle=-90]{rs3_2.ps}

\includegraphics[width=6.0cm,height=15cm,angle=-90]{fasa_2.ps}
\end{center}
\caption[example] {Lateral color diagrams for RS3 (top) 
             and FASA2 (bottom) ideal designs.}
\end{figure} 

\begin{figure}[h]
\begin{center}
\includegraphics[height=17cm]{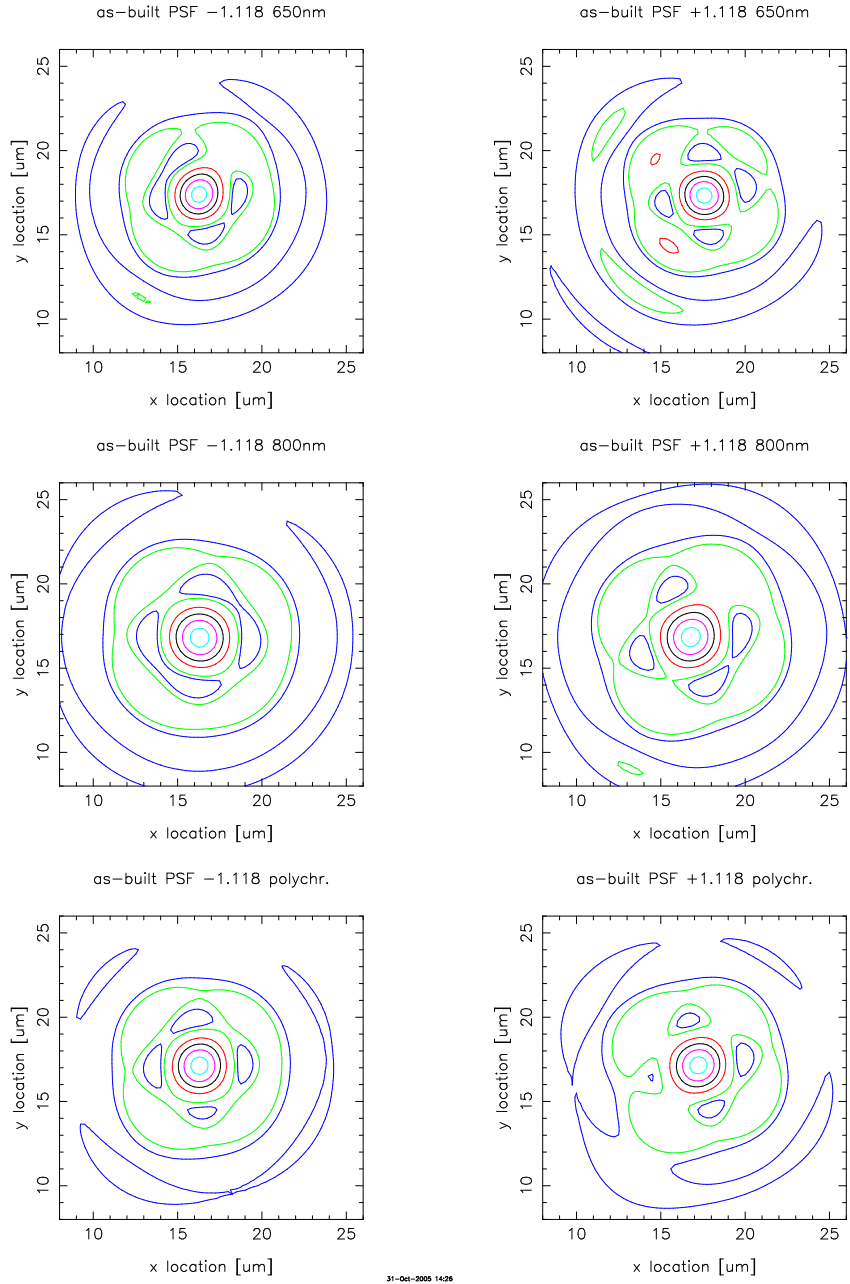}
\end{center}
\caption[example] {``as-built" PSFs contour plots of FASA2 design.}
\end{figure}

\section{Astrometric Performance}

\begin{table}[h]
\caption{Lateral color from ``as-built" FASA2 design PSFs 
         for sample points in the focal plane and extreme 
         monchromatic wavelengths (650$-$800 nm) as well
         as polychromatic (blue$-$red star, see text).
         Maximal position differences in $\mu$m are listed
         for original PSF and after folding with 0.8 arcsec 
         FWHM seeing, separately for the $x$ and $y$ axis.}
\label{tab:latcol}
\begin{center}       
\begin{tabular}{|l|cccc|cccc|} 
\hline
\rule[-1ex]{0pt}{3.5ex}        & \multicolumn{4}{c|}{monochromatic} 
                               & \multicolumn{4}{c|}{polychomatic} \\
\rule[-1ex]{0pt}{3.5ex} field  & \multicolumn{2}{c}{orig.~PSF}
                               & \multicolumn{2}{c|}{0.8" seeing}
                               & \multicolumn{2}{c}{orig.~PSF}
                               & \multicolumn{2}{c|}{0.8" seeing} \\
\rule[-1ex]{0pt}{3.5ex} point  &   $x$   &  $y$  &   $x$   &  $y$ 
                               &   $x$   &  $y$  &   $x$   &  $y$ \\ 
\hline
\rule[-1ex]{0pt}{3.5ex} on axis& 0.38& 0.54& 0.11& 0.16&  &  &  &  \\
\rule[-1ex]{0pt}{3.5ex}  -1.118& 0.03& 0.59& 0.20& 0.30&  &  &  &  \\
\rule[-1ex]{0pt}{3.5ex}  +1.118& 0.77& 0.49& 0.13& 0.01& 0.35& 0.23& 0.07& 0.01\\
\rule[-1ex]{0pt}{3.5ex}  -1.500& 0.17& 0.64& 0.20& 0.27&  &  &  &  \\
\rule[-1ex]{0pt}{3.5ex}  +1.500& 0.91& 0.49& 0.11& 0.01& 0.22& 0.15& 0.01& 0.01\\
\hline
\end{tabular}
\end{center}
\end{table}

Figure 8 shows the lateral color for the RS3 and FASA2 designs.
Both are acceptable for URAT.

For the FASA2 design ``as-built" point spread functions (PSFs)
were generated with the adopted tolerances and manufacturing
errors.
Figure 9 shows example contour plots of these PSFs at 2
worst-case locations in the focal plane (left, right), 
for 650 nm (top), 800 nm (middle) monochromatic, and 
flat-weighted polychromatic data (bottom).
The contour levels are 90\%, 70\%, 50\%, 30\%, 10\%, and 5\%.

After folding with a symmetric Gaussian function representing
realistic seeing conditions from 0.4 to 1.6 arcsec FWHM
the image profiles look even more symmetric.
For a quantitative analysis, image centroids were calculated
with the simple ``center-of-mass" algorithm cutting the
profile at various contour levels.
For lateral color investigations the maximal position differences
(in $\mu$m) are summarized in Table 4.
The PSF peaks are defined by the 90\% flux level and above.
On the left hand side the position differences for the
extreme, monochromatic colors (650 nm and 800 nm) are listed,
separately along the $x$ and $y$ axis (meridional and sagittal,
respectively). 
On the right side the differences between 2 
polychromatic PSFs are shown, with weighting
of 1,1,2,5 for 650, 700, 750 and 800 nm respectively
for a red star and weighting 5,2,1,1 for a blue star.

\begin{table}[h]
\caption{Position differences ($\mu$m) between extreme (90\% and 1\%)
         contour levels for the ``as-built" FASA2 design PSFs.
         Data are shown for the original PSF and after folding
         with 0.4, 0.8, and 1.6 arcsec seeing, separately for the
         $x$ and $y$ axis.}
\label{tab:coma}
\begin{center}       
\begin{tabular}{|l|c|cccc|cccc|} 
\hline
\rule[-1ex]{0pt}{3.5ex} field& & \multicolumn{4}{c|}{along x-axis} 
                               & \multicolumn{4}{c|}{along y-axis} \\
\rule[-1ex]{0pt}{3.5ex} point&color  & orig & 0.4 & 0.8 & 1.6  
                                     & orig & 0.4 & 0.8 & 1.6  \\ 
\hline
\rule[-1ex]{0pt}{3.5ex} on axis& 650&0.58&0.44&0.22&0.07& 0.95&0.68&0.31&0.05\\ 
\rule[-1ex]{0pt}{3.5ex}        & 800&0.25&0.25&0.17&0.10& 0.40&0.40&0.23&0.09\\
\rule[-1ex]{0pt}{3.5ex}        &poly&0.47&0.35&0.19&0.08& 0.80&0.53&0.38&0.07\\
\hline
\rule[-1ex]{0pt}{3.5ex} -1.118 &650 &0.10&0.17&0.17&0.10& 0.96&0.78&0.41&0.06\\
\rule[-1ex]{0pt}{3.5ex}        &800 &0.13&0.11&0.07&0.10& 0.40&0.38&0.22&0.11\\
\rule[-1ex]{0pt}{3.5ex}        &poly&0.05&0.10&0.10&0.10& 0.71&0.55&0.28&0.07\\
\hline
\rule[-1ex]{0pt}{3.5ex} +1.118 &650 &1.29&0.73&0.27&0.09& 0.96&0.64&0.29&0.05\\
\rule[-1ex]{0pt}{3.5ex}        &800 &0.28&0.30&0.28&0.16& 0.48&0.47&0.30&0.11\\
\rule[-1ex]{0pt}{3.5ex}        &poly&0.84&0.54&0.27&0.11& 0.78&0.58&0.29&0.07\\
\rule[-1ex]{0pt}{3.5ex}        &polB&1.06&0.61&0.28&0.10& 0.88&0.60&0.30&0.06\\
\rule[-1ex]{0pt}{3.5ex}        &polR&0.62&0.41&0.28&0.14& 0.64&0.53&0.29&0.09\\
\hline
\rule[-1ex]{0pt}{3.5ex} -1.500 &650 &0.04&0.24&0.22&0.09& 1.05&0.88&0.47&0.06\\
\rule[-1ex]{0pt}{3.5ex}        &850 &0.29&0.23&0.12&0.10& 0.42&0.38&0.23&0.11\\
\rule[-1ex]{0pt}{3.5ex}        &poly&0.17&0.19&0.14&0.10& 0.72&0.58&0.30&0.08\\
\hline
\rule[-1ex]{0pt}{3.5ex} +1.500 &650 &1.25&0.63&0.23&0.09& 0.99&0.67&0.31&0.05\\
\rule[-1ex]{0pt}{3.5ex}        &850 &0.10&0.17&0.23&0.17& 0.48&0.52&0.32&0.11\\
\rule[-1ex]{0pt}{3.5ex}        &poly&0.73&0.43&0.24&0.12& 0.81&0.61&0.31&0.08\\
\rule[-1ex]{0pt}{3.5ex}        &polB&0.67&0.41&0.24&0.12& 0.72&0.61&0.32&0.07\\
\rule[-1ex]{0pt}{3.5ex}        &polR&0.48&0.30&0.23&0.16& 0.64&0.57&0.32&0.10\\
\hline
\end{tabular}
\end{center}
\end{table}

\clearpage

Image symmetry has been quantified similarly using ``center-of-mass"
image positions derived at extreme (90\% and 1\%) contour levels.
A position derived from the peak location of an image profile would
correspond to an astrometric observation of a faint star, while
for a bright star the entire profile starting at a low level above
the background would be used for a position fit.
For symmetric images there would be no difference in position.
This is even true for astigmatic images with different profile
widths along 2 axes.
For asymmetric image profiles (coma aberration) however there is such 
a ``magnitude equation", a systematic error in star position as a 
function of the brightness of the star.
The dominant source of coma here comes from alignment errors.
Thus alignment tolerances are very tight for astrometric instruments.
Image asymmetry is typically not tolerable on a level where the
the optical design would still give acceptable performance as
judged by the Strehl ratio alone.

Table 5 lists the position differences ($\mu$m) between extreme
contour levels of the same PSF.
These numbers have been obtained from the original ``as-built"
FASA2 design PSFs as well as after folding with 0.4, 0.8, and 1.6
arcsec seeing. 
Again, results are presented separately for the $x$ and $y$ coordinate
(meridional and sagittal, respectively).
For each field point, monochromatic as well as polychromatic PSFs
are analyzed, with ``poly" meaning a flat-weighted polychromatic
PSF while ``polyB" and ``polyR" are for the aforementioned
blue and red stars, respectively.

For 0.4 arcsec seeing, position offsets between peak and low-level
contour centroids exceed the requirement, while results for
0.8 arcsec seeing and more are well below the 400 nm position
difference requirement.
Acceptable astrometric performance with respect to image
asymmetry is expected for about 0.7 arcsec seeing, which is sufficient
for the envisioned survey work of URAT.

Thermal and mechanical details of URAT are currently under
investigation.
A slight change of focus (scale) from exposure to exposure
is tolerable, however, higher order geometric distortions
as well as residual color and magnitude dependent systematic
errors need to be constant to high precision (0.1 $\mu$m level).
More details on the FASA2 design can be found elsewhere
in these proceedings \cite{fasa2}.

\end{document}